# TOWARDS AN ACCOUNT OF COMPLEMENTARITIES AND CONTEXT-DEPENDENCE

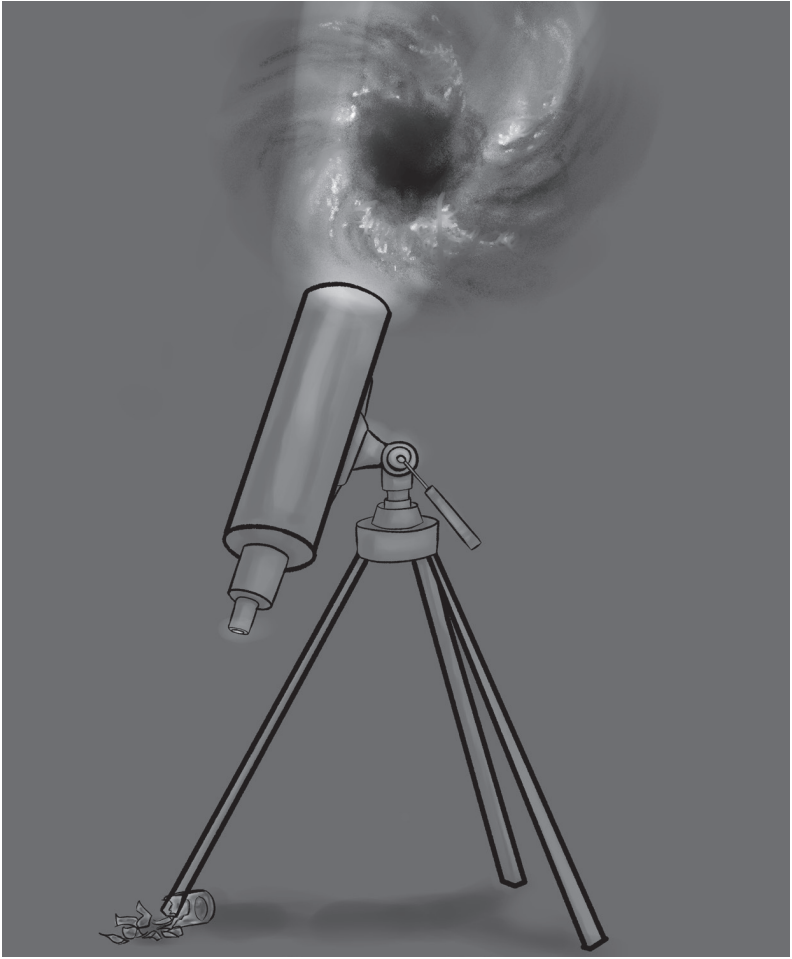

ABSTRACT

Modern physics proposals present deep tensions between seemingly contradictory descriptions of reality. Views of wave-particle duality, black hole complementarity, and the Unruh effect demand explanations that shift depending on how a system is observed. However, traditional models of scientific explanation impose a fixed structure that fails to account for varying observational contexts. This paper introduces context-dependent mapping, a framework that reorganizes physical laws into self-consistent subsets structured around what can actually be observed in a given context. By doing so, it provides a principled way to integrate complementarity into the philosophy of explanation.

HONG JOO RYOO

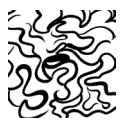
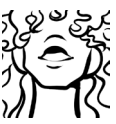

## I. INTRODUCTION

In the philosophy of science, explanation often involves deriving one set of propositions from another. The *deductive-nomological* (DN) account explains a desired event (*explanandum*) by deducing it from general laws through an *entailment structure*—a logical setup where certain premises necessarily lead to specific conclusions.[1] Lewis' account of causation focuses on causal histories within a system, while Phillip Kitcher's *Unificationism* seeks the most general explanations by deriving many conclusions from a small number of basic principles.[2] Michael Strevens' *Kairetic account* builds on these by incorporating entailment structures, causal relevance, and *Unificationism*-based principles.[3]

In this paper, I argue that current accounts (particularly the DN and *Kairetic accounts*) of explanation struggle to accommodate *complementarity*—cases where observed properties depend on the measurement context—because they impose fixed explanatory structures that lead to contradictions and misrepresentations. To resolve this, I develop *context-dependent mapping* ($f_c$), a new framework that partitions explanatory models based on the observational context.[4] Through examples like wave-particle duality and black hole information, I demonstrate how $f_c$ integrates complementarity into scientific explanation.

## II. THE KAIRETIC ACCOUNT

Strevens characterizes the *Kairetic account* as "fully causal," meaning that it relies entirely on causation to explain phenomena, excluding any non-causal or correlational elements.[5] The *Kairetic account* relies on a *causal model*: an entailment structure (as in DN) with causal premises. This model enables the identification of *difference-makers*: elements that, when altered, significantly change the outcome.

To identify difference-makers, Strevens uses *abstraction*: a process that refines a detailed causal model (M) into a simplified version (M*) that preserves only the necessary factors. For instance, explaining a cup breaking could start with a complex causal model that includes all physical forces and molecular dynamics. Through abstraction, this model simplifies to key conditions—like the cup being dropped on a hard surface—that are sufficient to explain the breakage. The *Kairetic account* isolates essential causal elements in a process similar to *Unificationism* but focuses on capturing relevant causal influences. The final, most abstract model contains only the core difference-makers.[6]

Finally, a *standalone explanation* is a fully developed causal model for an event (*e*) composed exclusively of elements that directly influence *e*—the difference-makers. This explanation is constructed from *explanatory kernels*: progressively abstracted causal models that maintain the causal requirements of prior models.[7]

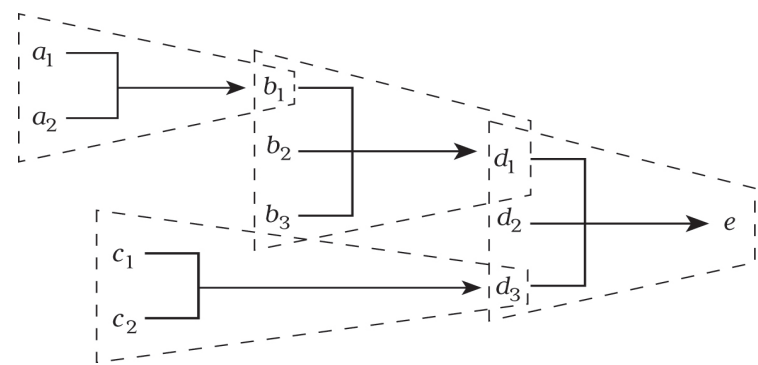

Figure 1: Sample Standalone Explanation[8]

Figure 1 showcases a typical standalone explanation constructed through four kernels (each grouped by dashed lines). The arrows represent causal entailment: for instance, in the first kernel, difference-makers $a_1$ and $a_2$ jointly entail $b_1$.[9] The first kernel in this chain isolates the essential causal factors, providing a focused explanation that captures all and only the components necessary for the event.

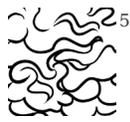
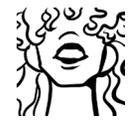

## III. DISJUNCTIONS AND MULTIPLE REALIZATIONS

In this section, I review two objections to the *Kairetic account* as a foundation for introducing the notions of similarity spaces and frameworking.

The first objection concerns disjunctions: overly abstract explanations may introduce unrelated causes, making them unclear. For example, explaining a cup shattering from a fall by either a cat bumping into it or an acidic chemical reaction leads to two unrelated causes.

To avoid this, Strevens argues that causal models should only include causes that operate under similar physical principles.[10] He formalizes this through the concept of similarity space, which is a collection of physical laws with a measure of how similar those laws are. By restricting an explanation to a single region of *similarity space*, we avoid combining unrelated causes and preserve explanatory clarity. In this case, we explain the cup's breakage using gravitational forces, without bringing in unrelated chemical laws.

On the other hand, the *multiple realizability objection* argues that high-level phenomena, like stress reduction, can arise from distinct mechanisms (e.g., exercise, therapy, medication), making it difficult to explain them with a single model.

Strevens resolves this concern through the practice of frameworking, which holds certain background conditions fixed to simplify explanations such as treating mental states as beliefs in psychology or assuming rational agents in economics.[11] Frameworking assumes one similarity space region while setting aside others, effectively taking for granted a specific explanatory perspective rather than addressing all realizations simultaneously.

## IV. PROBABILISTIC EXPLANATION

Quantum mechanics introduces challenges for the *Kairetic account*. In the current model of micro-scale physics, quantum mechanics is probabilistic. As Strevens puts it, "probabilistic explanations are the only possible explanations."[12] This probabilistic feature is genuinely irreducible to our current knowledge, classified as *simple probabilities*. Consequently, Strevens declares the *Kairetic account* does not have "much that is novel to say" about it.[13]

Ultimately, Strevens directs the audience to Railton's *Deductive-Nomological-Probabilistic* (DNP) account, which is simply the DN account's entailment structure with probabilistic laws.[14] Indeed, Strevens affirms that "[the *Kairetic account* for simple probabilities] follows the broad outlines of the DNP and similar accounts."[15] If one were to attempt to create a standalone explanation, especially involving quantum mechanics, then it is fitting to use a structure similar to that of the DNP account to construct the kernels that arise from quantum mechanics. Strevens depicts such an explanation of Rasputin's death.

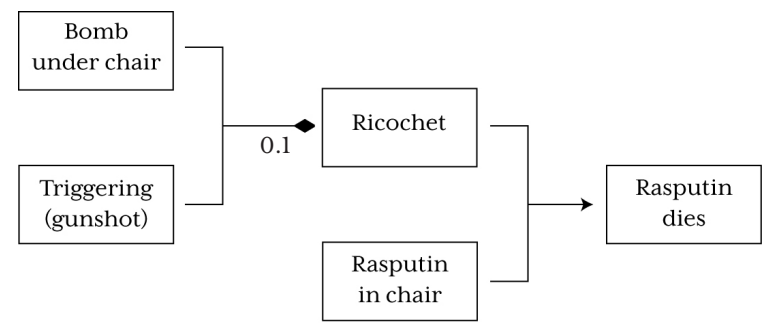

Figure 2: Sample Probabilistic Standalone Explanation[16]

The event involves a bomb under Rasputin's chair and a gunshot that is intended to trigger the bomb. With a small probability, represented by 0.1, the bullet does not trigger the bomb and instead causes a ricochet into Rasputin. Thus, DNP models form the basis of a standalone explanation incorporating the ricochet's probabilistic elements. Strevens claims: "a deep standalone explanation must spell out the properties of the fundamental-level laws in virtue of which the explanandum has the physical probability it does."[17]

Strevens' thought regarding the DNP account is that it serves to explain a simple probability kernel which can be a part of a standalone explanation in the *Kairetic account*. The entailment structure of the DNP account allows one to construct a standalone explanation like the one in Figure 2. Since both the *Kairetic* and DNP accounts rely on physical laws within a given similarity space, it is crucial to examine how they handle complementarity. The following sections explore various complementarities and how these explanatory frameworks accommodate—or fail to accommodate—context-dependent observations.

---

10   This method is formally named "cohesion." See Strevens, "Causal and Unification," 17.

11   Strevens, *Depth*, 154–59.

12   Strevens, *Depth*, 341–58.

13   Strevens, *Depth*, 341–58.

14   Peter Railton, "A Deductive-Nomological Model of Probabilistic Explanation," *Philosophy of Science* 45, no. 2 (University of Chicago Press, 1978): 206–26, 13.

15   Strevens, *Depth*, 358.

16   Strevens, *Depth*, 442.

17   Strevens, *Depth*, 362.

## V. A BRIEF INTRODUCTION TO COMPLEMENTARITIES

While Strevens asserts the probabilistic nature of quantum mechanics as simple probabilities, some may advocate for a view of complementarities that is not completely probabilistic but unequivocally dependent on the apparatus (reference frame, experimental instruments, and other entities in a system) used in observation.[18] The *complementarity principle* posits that entities/systems can display apparently contradictory properties depending on the way they are observed or measured. These manifested properties are mutually exclusive yet equally necessary for a consistent description of the system with our physical laws. In this section, I illustrate this perspective through key examples.

The *wave-particle duality*, for instance, requires the apparatus to be the sole determinant of the manifested features. In the double-slit experiment, light manifests itself as a wave and produces interference patterns. However, in the case of the photoelectric effect—where light is shined onto a metal—light behaves as a particle, which scatters and interacts with the metal as though it in itself is an energy-carrying particle.[19] Depending on the experimental context, light appears in different forms. The identity of a particle is two-fold: it is a particle and a wave in various contexts.

The nature of the observation of the wave-particle duality (whether it reveals wave or particle behavior) is not probabilistic in the sense of a quantum superposition. When used, the experimental apparatus instead unequivocally influences the nature of the observed phenomena: Depending on the experiment, there seems to be a set of relevant physical laws corresponding to waves or particles. Beyond this complementarity, there are plenty of paradoxes and complementarities that arise from our physical theories.

Furthermore, Stephen W. Hawking's radiation affirms that when an object enters a black hole (crosses a radius called the event horizon), we expect to see the object emitted as radiation.[20] The *information paradox* goes as such: Quantum mechanics predicts that the radiation will contain all of the relevant information of the object that entered the black hole, but general relativity suggests that this information will be destroyed in the sense of the information not being able to escape the event horizon.[21]

*Black hole complementarity* is then a proposal to remove the frame-independence of this event.[22] Albert Einstein's equivalence principle states that an object that crosses the event horizon will continue to experience falling toward the center of the black hole rather than just be emitted as radiation right away. This forms the crux of the complementarity: If we are in the reference frame of the falling object, we would continue to fall, but if we were on a reference frame outside of the event horizon, then we would see an object enter and then exit as radiation. In modern physics, complementarities are often proposed as solutions to disagreements between theoretical frameworks.

Lastly, the geometry of space-time and reference frames impact the observed quantum effect: The Unruh Effect described by Luis C. B. Crispino, Atsushi Higuchi, and George E. A. Matsas, suggests that an observer in acceleration will perceive a warm bath of radiation, even if they are in what an inertial observer would describe as a perfect vacuum, void of quantum states.[23] This effect is a manifestation of the observer-dependent nature of the vacuum state in quantum field theory.

Again, a complementarity is proposed for this paradox, referred to as the complementarity principle for the Rindler horizon.[24]

---

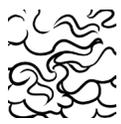
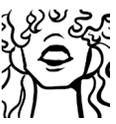

This principle suggests a relativity of observational knowledge for an accelerating observer in an analogous manner to black hole complementarity. Just as in the wave-particle duality case, there seems to be a set of relevant physical laws corresponding to each complementarity feature. The ultimate aim of this paper is to accommodate complementarities into the DNP and *Kairetic accounts*.

Thus far, I have reviewed the following points:

1. Strevens points towards the DNP account for simple probabilities within a standalone explanation of the *Kairetic account*.

2. There are paradoxes in our theories (wave-particle, information, Unruh effect, etc.).

3. Complementarities, as proposed solutions to those paradoxes, inherently depend on context.

The thoughts developed in the following sections will consist of the following:

4. If one were to use frameworking in the manner described by Strevens, then the complementarity view is misrepresented.

5. If one were to use the DNP account, then contradictions may arise.

6. Using point 3, it is possible to construct a mapping $f_c$ that partitions a similarity space containing contradictory laws into multiple *self-consistent* similarity subspaces: regions of similarity space that do not contain contradictions.

7. If one wishes to explore an account of explanation involving complementarities, then one can utilize the mapping in point 6.

## VI. IN THE FACE OF COMPLEMENTARITIES

Given that Strevens concedes the *Kairetic account* does not address simple probabilities, I revisit frameworking to show why it misrepresents complementarity in general (point 4) and set the stage for an alternative way to partition similarity spaces. Recall that Strevens defines similarity space as encompassing all physical laws, grouping causal models by shared physical principles. It follows that dual phenomena (e.g., wave-particle duality) occupy two separate, incohesive regions of this space. As discussed earlier (point 3), wave-like and particle-like behaviors are two facets of the same underlying reality; the aim is to express the potential to manifest different properties under different conditions. Frameworking, however, imposes an artificial divide, treating one region of similarity space (e.g., particle-like behavior) as fundamental while excluding the other (e.g., wave-like behavior) within the same explanatory model. This misrepresents complementarity, which requires both descriptions to be context-dependent and simultaneous, rather than rigidly and classically distinct.

Moreover, one can argue for point 5 by demonstrating that the laws themselves are contradictory. This was briefly done in our discussion of paradoxes (point 2). Without frameworking, if one were to utilize the DNP account's entailment structure, then our physical laws regarding the wave-like and particle-like features will produce contradictions as both features cannot hold at once: only one manifests itself. Similarly, when involving relativistic laws, the entailment structure of the DNP account would lead us to the paradoxes involving the Unruh effect or the black hole information paradox, as we would have conflicting theories as the premises.

Accordingly, until a theory of unification reconciles quantum mechanics and general relativity, there are inevitable contradictions within the entailment structure of the DNP account. To resolve this, I propose a framework that dynamically assigns self-consistent sets of physical laws based on observational context. The next section introduces context-dependent mapping $f_c$, which provides a systematic method for this assignment.

## VII. $f_C$: CONTEXT-DEPENDENCE AS A REPRESENTATON OF COMPLEMENTARITIES

To demonstrate point 6, I define $f_c$ as a mapping from a set of contexts to an appropriate set of physical laws. I first categorize observations into *context space*, which is a set of phenomena that we know involve paradoxes.[25] Within this space, I identify *context subsets*, denoted as $c_i$, where each subset consists of phenomena that exhibit a single complementarity feature under specific experimental conditions.

---

25  Context space parallels similarity space: it groups physical phenomena as similarity space groups physical laws. For motivation on similarity space, see Strevens, *Depth*, 104–05.

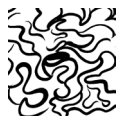
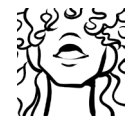

Taking the double slit case as an example, let light manifest as a wave. This physical setup involving the slit and light will be a part of a context subset $c_{Wave}$ alongside all other possible setups manifesting wave features of quantum entities. More examples of $c_{Wave}$ involve multiple slits, optical setups, etc. Setups involving accelerating reference frames producing the Unruh Effect will also be a part of a context subset $c_{UA}$ and all setups that can be treated as a black hole's matter system (the frame which enters the event horizon) will be in $c_{BHM}$.

The next step is to identify a mapping $f_c$ from a context $c_i$ onto a *similarity subspace* ($S_i$): a particular region of similarity space with non-contradictory laws. Namely, there will be no major overlap of general relativity and quantum mechanics, no overlap of wave and particle features of quantum entities, etc. For instance, if the context involves a double slit, then the particle is to be considered in terms of wave formalism $S_{Wave}$, and if there is scattering, then it is to be considered in terms of particle formalism $S_{Particle}$. Through this partition, $f_c$ ensures that only the relevant laws for the observed complementary feature are applied.

Using empirical information from past experiments and verified theories, one can assign the relevant laws to a physical system. The thought is to have the mapping be grounded in empirical data and theoretical frameworks that are accepted independently of the specific explanations being constructed. The mapping aims to, depending on the context/setup, utilize disjoint and non-contradictory laws for the entailment structure. Figure 3 depicts a sample mapping.

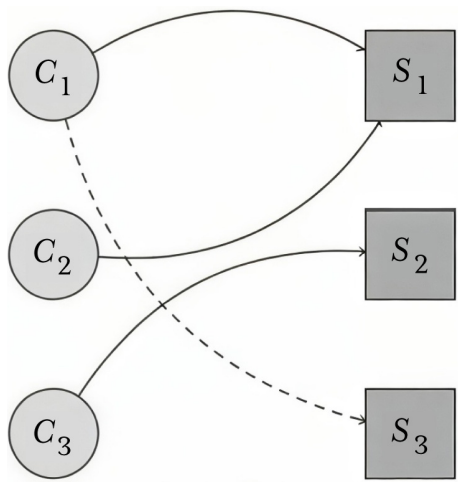

Figure 3: Visual Representation of $f_c$

In this figure, $c_i$ represents the set of contexts that manifest one complementary feature: this may be the set of all accelerating frames in the Unruh effect or the set of all matter frames in the presence of a black hole. $S_i$ represents the consistent similarity subspaces of similarity space. This may be the equations governing quantum field theory, black hole relativity, wave physics in quantum mechanics, etc.

This section demonstrates point six. The mapping $f_c$ is intended to be an accommodation of complementarities—of a context $c_i$ unequivocally influencing the relevant set of laws $S_i$. Its navigation around the contradictions of similarity space further echo complementarities as a proposed solution to physical paradoxes. If one desires to apply this mapping, in the case of dual phenomena, one can consider each mapping onto different similarity subspaces and use the corresponding physical principles to construct a DNP or similar account of scientific explanation.

## VIII. EXAMPLE OF THE MULTI-SLIT MESH

In this section, I will aim to further establish that an account of complementarities can be constructed through $f_c$ mapping (point 7). I will construct an example case to demonstrate how one could utilize context-dependent mapping. Consider an interference phenomenon arising from shining a light towards multiple slits in a mesh (with sufficient geometry such that diffraction occurs). From our physics, in this context, the laws that apply are those involving wave-like manifestations of light. Figure 4 depicts the complementarity's relevant similarity subspaces and the sets of contexts to which each one corresponds.

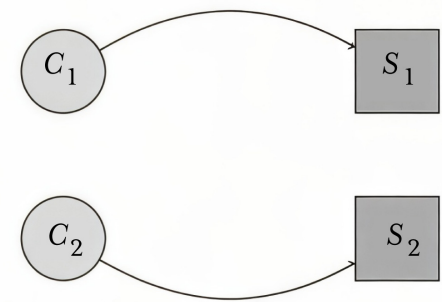

Figure 4: Complementarity of Wave-Particle Features

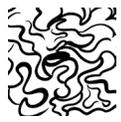
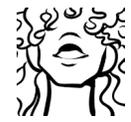

In this figure, the set of contexts $c_i$ contains the contexts that involve the manifestation of light as a wave, such as cases with double (or even multiple) slits with appropriate length(s). It is a set of all such contexts (of the energy of surrounding particles, the reference frames, and other physical background conditions) for which one may identify wave-like features of particles. Accordingly, $c_2$ involves the contexts for which light manifests itself as a particle such as scattering, photoelectric effect, or near photodetectors. $S_1$ and $S_2$ are the similarity subspaces that involve the physics of a wave and the physics of a particle. In the case of the mesh, one with knowledge of physics would make these judgments:

1) The set of contexts $c_i$ is appropriate to use.

2) $f_c$ maps $c_i$ onto $S_i$.

3) Using the laws in $S_i$ we can construct an entailment structure.

The first judgment arises from the observation of the explanandum (of interference), and the second judgment comes from the physics background: having knowledge of the dual phenomena allows one to see that it is fitting to utilize wave-like laws of quantum mechanics. I will now continue to briefly construct a structure of the DNP account using $S_i$ laws. The model would look like this: {the particles' wavefunction from the Schrodinger Equation, laws of constructive and destructive interference of waves, etc.} and these would entail the explanandum: the interference pattern in the mesh.

## IX. EXAMPLE OF THE OPTICAL TELESCOPE

I offer an example in which the interference patterns of quantum mechanics are considered through the DNP account within a standalone explanation of the *Kairetic account*. Suppose one wants to explain the phenomena of a telescope (which uses a mesh similar to the previous example) showing a certain spectrum of light from a star.[26] The explanandum, in this case, is the telescope reading, which takes the form of a spectrum shown on a screen. If one wishes to use the mapping, then one can adopt the one in the previous example:

the light emitted from the star and the diffraction mesh geometry $c_i$ determine that the similarity subspace from which the laws are drawn would be the same $S_1$. Using the laws in $S_1$, one can use the DNP account to demonstrate that, in this case of the complementarity, we have wave-like features within the telescope and a certain interference pattern.

One can go on to consider the *Kairetic account*. A causal model M can look like this: {we have the aforementioned quantum interference pattern (explained by the DNP account), the other classical optical properties of the telescope, a spectrometer that indicates calcium if the pattern corresponds to a certain family of other patterns, etc., if these previous elements are satisfied then we get the spectrum displayed in the telescope}. The model would then continuously be abstracted so that the only remaining elements are difference makers. Figure 5 depicts the standalone explanation as a parallel to the diagram of Rasputin's death—the kernel of the Interference Pattern is then explained by the first example (DNP structure) for which the mapping onto the wave (interference) laws has been performed.

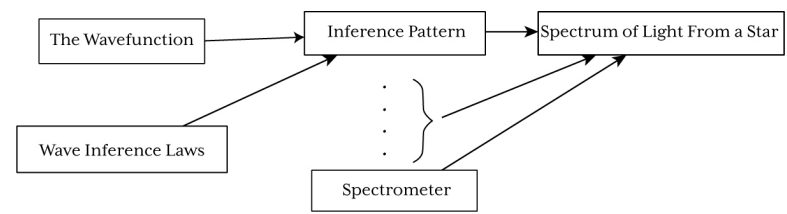

Figure 5: Optical Telescope Standalone Explanation

Using the DNP account and specifying the laws within it, it is possible to explain the quantum mechanical aspects of a telescope. The DNP account explains the probabilistic nature of photon interaction at the quantum level, while the *Kairetic account* is used to model how these interactions causally lead to reliable spectral readings under specific observational settings. This aligns with Strevens' thoughts about the DNP account being able to account for the simple probabilities. When constructing a standalone explanation, it may come to one's attention that the quantum interference pattern, which emerges within a model, would be the kernel of a DNP entailment analogous to the ricochet in Figure 2. The mapping, when used in the *Kairetic account*, may serve to give a causal model that does not sprout from a potentially contradictory set of laws but rather from a view of complementarity.

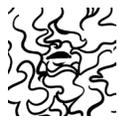
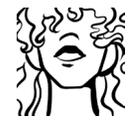

## X. CONCLUSION

In this paper, I offer a preliminary yet practical tool to integrate complementarity within the DNP and *Kairetic accounts*. The prescription goes as so: (1) identify the context $c_i$, (2) use $f_c$ to map onto a set of consistent laws $S_i$, (3) use those laws to construct an entailment structure, then (4) complete a standalone explanation. By mapping observational contexts to distinct similarity subspaces, context-dependent mapping $f_c$ offers a systematic navigation around the contradictions that arise from physics, representing complementarities. Further development of this thought may be particularly useful in cases of non-negligible quantum and relativity effects before an established unification.

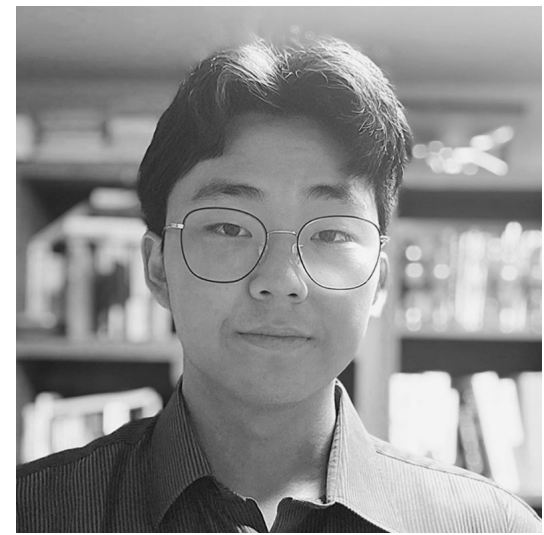

Hong Joo Ryoo earned a quadruple bachelor's degree in physics, philosophy, cognitive science, and applied mathematics from the University of California, Berkeley. His interests span philosophical physics, epistemology, metaphysics, and cognition. He will pursue a PhD in physics at Johns Hopkins University, aiming to complete a concurrent MA in philosophy.

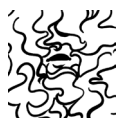
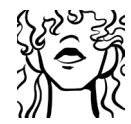